# Chapter 1

# Analysis of sensors for movement analysis

Marcos Faundez-Zanuy[0000-0003-0605-1282], Anna Faura-Pujol [1], Hector Montalvo-Ruiz[1], Alexia Losada-Fors [1], Pablo Genovese [1], Pilar Sanz-Cartagena [2]

[1] Fundació Tecnocampus, Avda. Ernest Lluch 32, Mataro, Spain
[2] Consorci Sanitari del Maresme, Mataro, Spain

e-mail: faundez@tecnocampus.cat

**Abstract.** In this paper we analyze and compare different movement sensors: micro-chip gesture-ID, leap motion, noitom mocap, and specially developed sensor for tapping and foot motion analysis. The main goal is to evaluate the accu-racy of measurements provided by the sensors. This study presents rele-vance, for instance, in tremor/Parkinson disease analysis as well as no touch mechanisms for activation and control of devices. This scenario is especially interesting in COVID-19 scenario. Removing the need to touch a surface, the risk of contagion is reduced.

## 1.1 Introduction

In the last years there has been an increment of 3D sensors for hand movement analysis, especially in the videogame industry. We forecast an increase on commercial models and applications due to COVID-19 issue. Another important application field is the analysis of Parkinson disease and essential tremor.

## 1.2 Hand movement analysis

Hand movement analysis detection can be a key aspect to avoid surface contact when trying to activate or select options in electronic systems. While for some ap-plications the sensor calibration is not an issue as only a coarse precision can be enough, this is not the case when dealing with medical applications. The purpose of this paper is to try to find the best product available on the market to make measurements of hand movements in 3D. We will calibrate the devices in order to know the optimal working range as well as the accuracy on the movement measurements.

One of the possible applications is hand tremor analysis for Parkinson disease, as alternative to wearing rings, such as [1]. In the following sections we summarize the main properties of the analyzed devices.

### 1.2.1 Microchip DV102014

This device is based on a Microchip technology called GestIC (http://ww1.microchip.com/downloads/en/DeviceDoc/40001716C.pdf). A 3D GestIC® sensor is the combination of a gesture controller (MGC3XXX) and a set of sensor electrodes. The GestIC electrodes consist of:
- 4 or 5 Receive electrodes (Rx) connected to Rx 0-4 pins of MGC3XXX
- 1 Transmit electrode (Tx)
- Isolation between Rx and Tx

Rx and Tx are made of any conductive material such as copper, metal mesh, indium tin oxide (ITO) or similar. The isolation between the electrodes can be any material which is non-conductive (PCB, glass, PET, etc.). An optional cover layer on top of the electrode must be non-conductive as well. Fig. 1 shows its physical aspect.

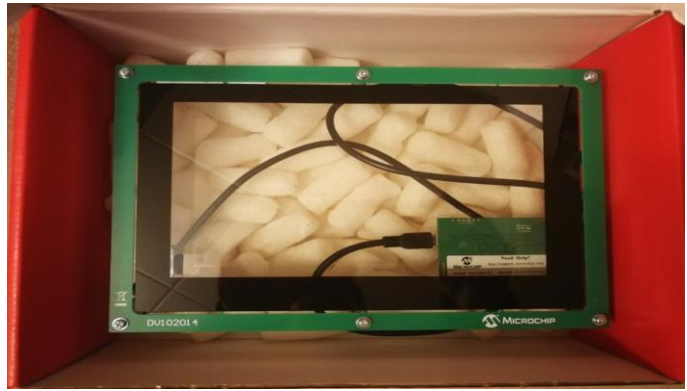

**Fig. 1.** Microchip DV102014 device in its package.

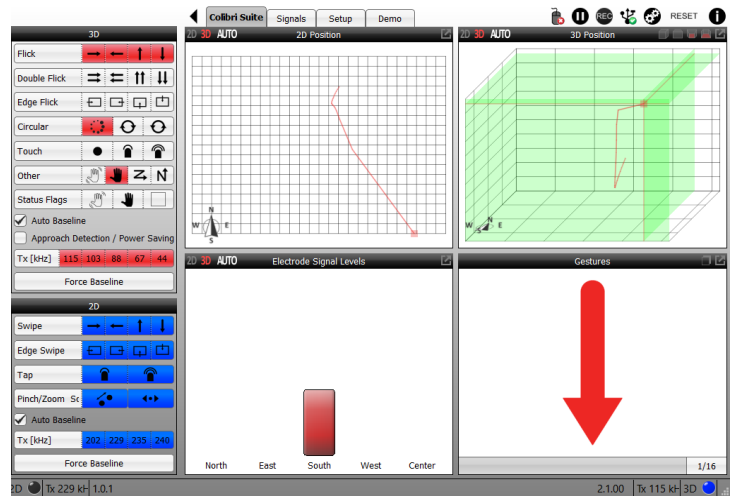

**Fig. 2.** Aurea software snapshot.

The device includes a software (Aurea) that can detect some specific movements and permits the recording of 3D movements (see Fig. 2).

There is also the possibility to use a SDK for programming specific applications. However, we experienced problems with the SDK, which was discontinued very quickly form the market and we had to discard this sensor.

Another important issue for a desktop scenario is that there are interferences from the bottom side of the sensor. When a user sits down in a chair and operates the sensor on a table it detects hand movements from the top but too movements from the legs under the table. This discouraged us from using this sensor.

### 1.2.2 Leap Motion

The second contact-less analyzed sensor is more compact the other two. The de-vice consists of two cameras and three infrared LEDs. These track infrared light with a wavelength of 850 nanometers, which is outside the visible light spectrum. The Leap Motion Controller's viewing range is limited to roughly 60 cm above the de-vice. This range is limited by LED light propagation through space, since it becomes much harder to infer the hand's position in 3D beyond a certain distance. LED light intensity is ultimately limited by the maximum current that can be drawn over the USB connection. Fig. 3 shows its scheme.

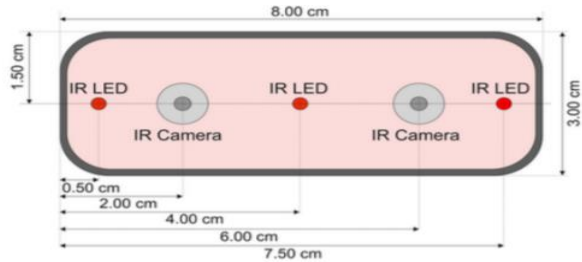

**Fig. 3.** Leap motion scheme.

There is also an SDK available for this device.
One of the published applications of this device outside the videogame / entertainment industry, is Parkinson Disease analysis [2]. However, to the best of our knowledge, its accuracy for physical measurements has not been tested before.
One important drawback is that the sensor can experiment problems to identify all the fingers. Especially when fingers touching each other, folded over the hand, or hidden from the camera viewpoint. This problem is solved with the next sensor, which is based on accelerometers and not on video images.

### 1.2.3 Noitom perception mocap

The third system is composed of interchangeable neuron sensors (inertial trackers) connected to a Hub. The Hub connects wirelessly to a computer through WIFI or can be wired directly through a USB connection. The system is powered by any external USB power pack. Fig. 4 shows the glove with sensors. Although this mocap can track the whole body, we just used the glove for the hand. Thus, one drawback when compared with previous systems is that a glove plenty of sensors must be put on the hand. This could limit some practical applications and, in some sense, is simi-lar to the wearing ring presented before [1], and cannot be considered a contactless device.
This system also provides an SDK but it was not necessary in our case to use it as data exportation in standard formats is quite straight forward. It can provide 60 fps (when using 18 - 32 Neurons) or 120 fps (with 17 Neurons or less).

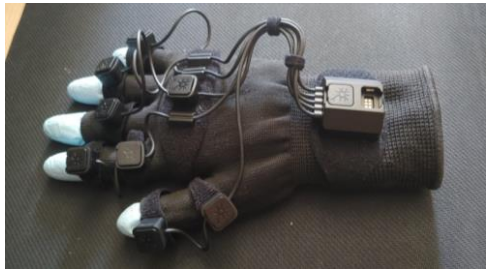

**Fig. 4.** Noitom perception mocap globe.

## 1.2.4 Android app for tapping analysis

The fourth possibility, although it is a contact sensor, it is an Android app based on detecting finger tapping on the own user's mobile so it is not a general sensor used by different people. It can be seen as a "personal sensor". Figure 5 shows a snapshot of the app. The user must touch the circle depicted on the surface as many times as possible during 15 seconds (a time counter appears on the right upper corner). The recorded information is sent by e-mail to a predefined e-mail account and later on, the regularity of the tapping can be analyzed.

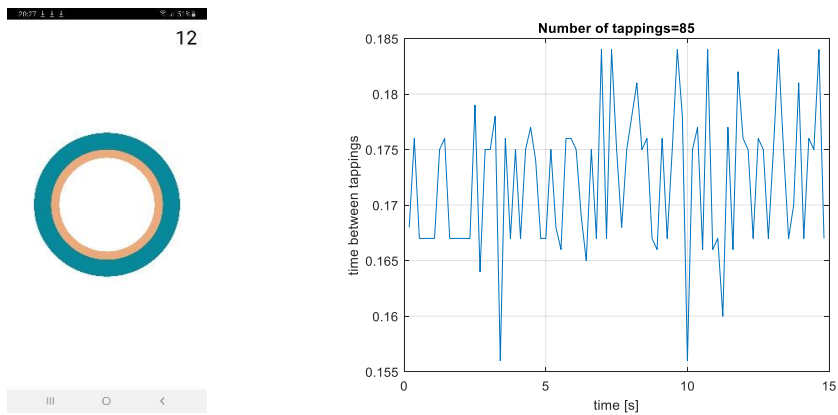

**Fig. 5.** Android tapping app (left) and example of registered signal (right).

## 1.3 Experimental results of hand movement analysis

### 1.3.1 Experimental setup for Leap Motion and Noitom Mocap

In order to obtain replicable experiments, we used a synthetic wood hand which permits finger movements and we mounted it in a tripod as shown in Fig. 6. In order to simultaneously acquire Noitom mocap signal and leap motion we put the Noitom mocap as a glove on the wood hand and used their softwares simultaneously. We performed static analysis at different heights as well as dynamic analysis by moving the tripod in a defined excursion range of several centimeters. The process was repeated for several ranges.
To obtain the real height ("gold standard") h provided by the tripod with the height obtained in leap motion we must apply the equation:
$$h\ [cm] = y - 19.55 - 1.09 \tag{1}$$
Where 19.55 cm is the distance between tripod scale and hand palm. 1.09 cm is the height of the leap motion device and y is the measurement obtained in the tripod scale.

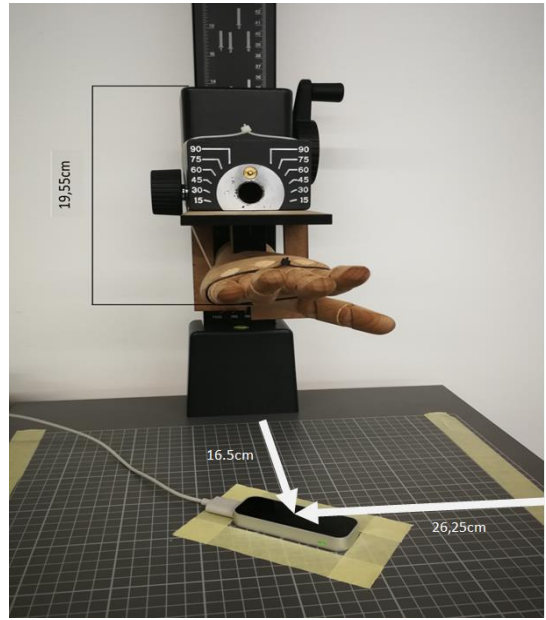

**Fig. 6.** Experimental setup (tripod, hand and leap motion).

### 1.3.2 Static analysis results

In a first experiment we acquired 50 consecutive samples at different levels. While expected result should be a fully flat line we obtain some minor oscillations (see Fig. 7). Fig. 8 compares the obtained measurement with leap motion with the ground truth. We can observe an underestimate for heights higher than 50 cm as well as overestimated height for distances smaller than 50 cm. The blue bars indicate the standard deviation of the measurements in each height.

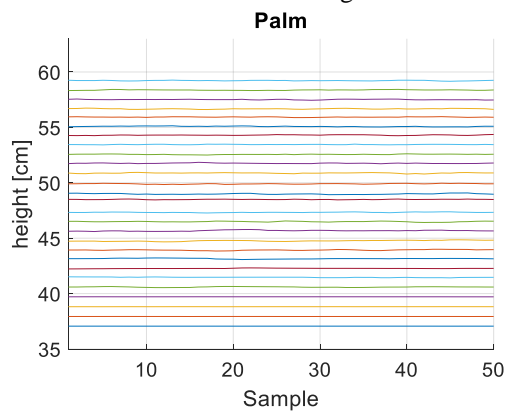

**Fig. 7.** Static measurements at different heights.

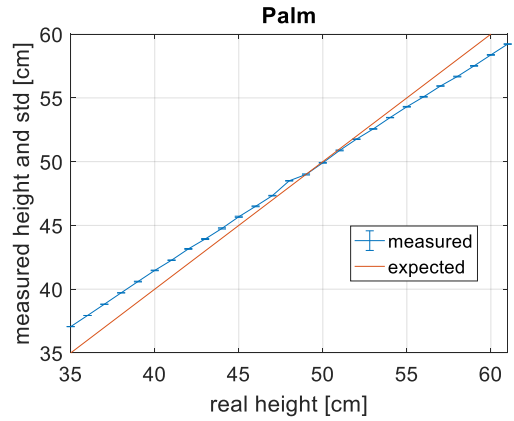

**Fig. 8.** Static measurements averages at different heights.

### 1.3.3 Dynamic analysis results

In a second experiment we moved the hand in a periodic movement and obtained the recorded distances shown in figure 9. We can observe that Noitom mocap detects larger excursions than leap-motion but basically the shape is similar.

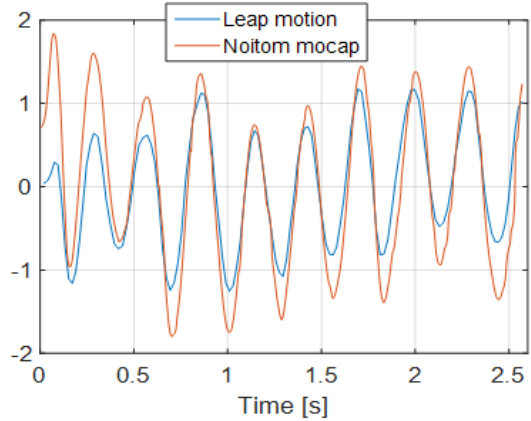

**Fig. 9.** Simultaneous dynamic measurements with leap motion and noitom mocap.

## 1.4 Foot sensor design and foot movement analysis

Foot movement detection is also an interesting field of analysis, especially when it comes to Parkinson diagnosis. Obviously, the range of applications is smaller than hand motion analysis but it has interest in medical applications.

The purpose of the new designed device is to register data during two exercises that are performed by Parkinson patients in a follow-up visit (see Fig. 10). Particularly, it seeks to quantitatively measure the range of motion, its speed and frequency, in these exercises that are performed under the supervision of a neurologist. This data well be used to support diagnose, to prescribe the medication dose, etc.

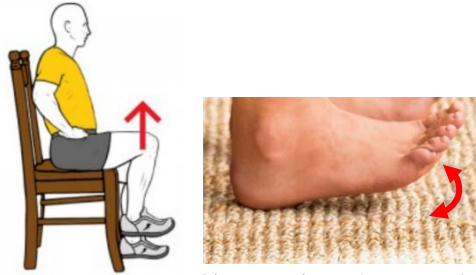

**Fig. 10.** Exercises performed by Parkinson patients that we seek to parametrize.

To select the best alternative for this purpose, we have analyzed different possibilities:
- A mobile app (AccDataRec) that records movement and can register foot motion when the phone is placed on top of this part of the body.
- A self-designed prototype that can either use laser or ultrasound technology.
- A system that includes several cameras, which is bought together with a software for image processing.

All options have been analyzed with a numeric assessment. This considered several aspects including how the motion capture was registered, the system's detection and frequency range and its precision. In the light of these, the best alternative is the self-designed prototype, which will be described next.

## 1.5 Experimental results of foot movement analysis

### 1.5.1 Experimental setup

We have deigned a prototype that takes advantage of the ultrasound technology. This prototype can be seen in Fig. 11. It consists of a DM board that has a hole to enable the proper transmission of the ultrasonic waves, which is attached to a wood-en box. All of it is fixed on top of a platform that avoids extra degrees of freedom on the movement. Inside the box, there is an ultrasound HC-SR04 sensor connected to an Arduino DUE board, which is linked to a PC.

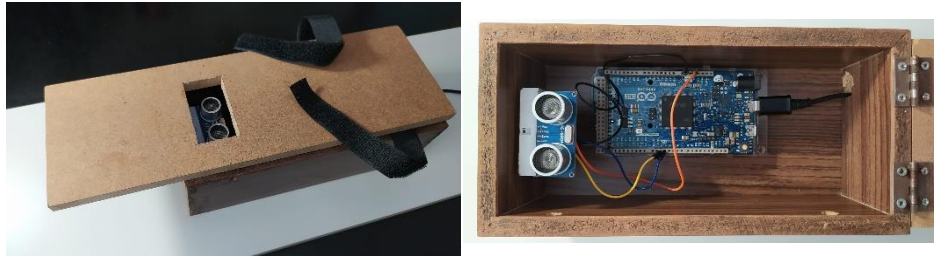

**Fig. 11.** Ad-hoc prototype.

For the parametrization of the first exercise shown in Fig. 10, the user will move his/her foot up and down, over the DM board (see Fig. 12). However, for the second exercise that can be seen in Fig.10, the patient will fasten his/her foot on top of the DM board with a black Velcro strap (see Fig. 13).

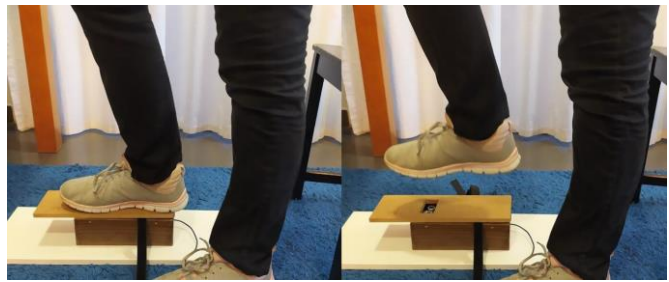

**Fig. 12.** First exercise performed by a user.

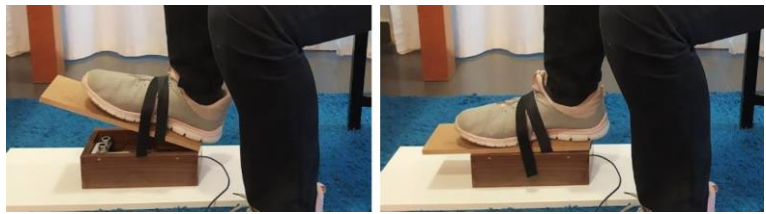

**Fig. 13.** Second exercise performed by a user.

### 1.5.2 Results

For the data registration we have used an Excel complement called Excel Data Streamer. The results show the range of motion of the user's movement (which will also be referred to as "distance" in the following sections), as well as its speed and frequency, either graphically and numerically.

### 1.5.2.1 Results on a healthy user

This system has been put into practice with several healthy users. The following figure show the results that have been obtained in one of these experiments, specifically when performing the second exercise.

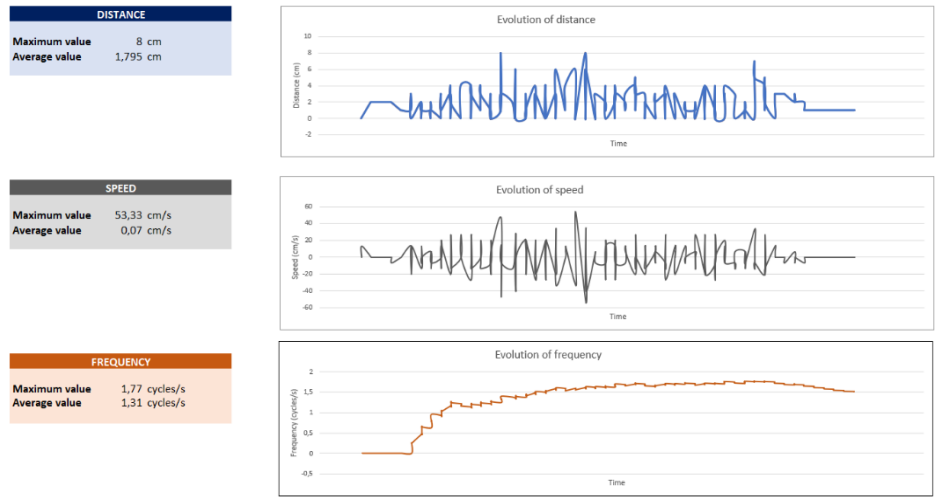

**Fig. 14.** Results on a healthy user.

### 1.5.2.2 Results on a Parkinson user

We have taken the system to Mataro's public Hospital, where it has been tried with several patients suffering from Parkinson disease.

In this case, some very interesting results have been seen. Whereas with healthy users it makes no difference to parametrize their right or left foot, with these collective it does. As it can be seen in the following figures, the obtained results differ if the user performs the exercise with their right or left foot. According to the hospital's neurologist, it is due to the fact that this disease always affects one half of the body more that the other. This prototype evidences and quantifies the difference between both.

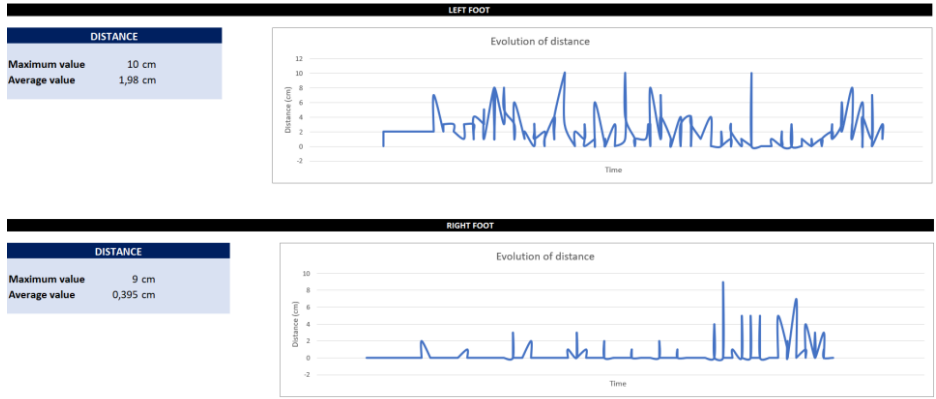

**Fig. 15.** Results on a Parkinson user (distance).

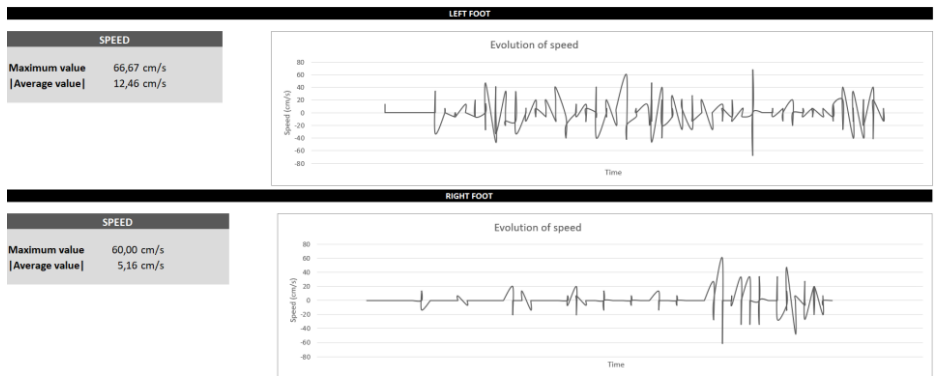

**Fig. 16.** Results on a Parkinson user (speed).

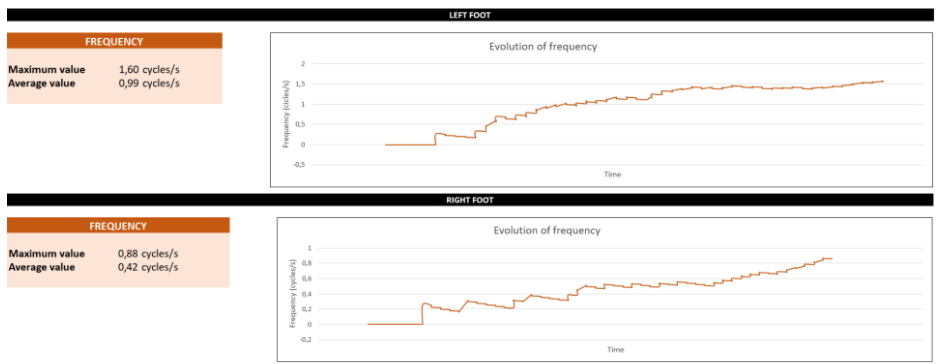

**Fig. 17.** Results on a Parkinson user (frequency).

In the light of the above, we can see that this patient's right half of the body is more affected by the disease than the left one. While the left foot presents results that are not far from the healthy users', the right foot's graphs show many more strokes when performing the same exercise.

## 1.6 Conclusions

For hand motion analysis we have evaluated three satisfactory devices:
- Leap motion, as a contactless device
- Noitom mocap, which can be considered as contact device, as all the us-ers must put the same glove on
- Android App to detect tapping, which can be considered a contact device, but each user hits its own mobile phone.

These three devices can be used for health applications and the leap motion for some other daily-life applications as it does not require any physical contact.

When it comes to the analysis of foot movement, we have proved that the de-signed prototype is totally capable of capturing motion data.

A future improvement of this system in the field of Parkinson could be to configure an application that the neurologist can install on his/her computer, and that al-lows him/her to record the data of all patients. Not only wouldn't the patient's data be compromised, but the application would also allow to create a file for each patient, so that their evolution is kept. Being able to see the evolution of each user, even while comparing the right and left side performance, allows experts to draw conclusions about the adjustment and effect of patients' medication, and the degeneration of this disease. All in all, a clear improvement in the diagnosis of Parkinson.

## Acknowledgement

This work has been funded by Spanish grant Ministerio de ciencia e innovación PID2020-113242RB-I00.